\documentclass[%
 reprint,
 superscriptaddress,
 amsmath,
 amssymb,
 aps,
 pra,
]{revtex4-1}

\usepackage{xcolor}
\usepackage{graphicx}
\usepackage{dcolumn}
\usepackage{bm}

\usepackage{hyperref}
\hypersetup{
	colorlinks = true,
	urlcolor   = blue,
	linkcolor  = blue,
	citecolor  = blue
}

\graphicspath{{./pics/}}
\newcommand{\beqn}{\begin{eqnarray}}
\newcommand{\eeqn}{\end{eqnarray}}
\newcommand{\beq}{\begin{equation}}
\newcommand{\eeq}{\end{equation}}

\newcommand{\cM}{{\cal M}}

\newcommand{\bra}[1]{|#1\rangle}
\newcommand{\ket}[1]{\langle #1|}

\newcommand{\myrule}{\rule[-5pt]{0pt}{16pt}}
\newcommand{\create}[1]{#1^\dagger}
\begin{document}
\preprint{APS/123-QED}
\title{Improved Heralded Schemes to Generate Entangled States From Single Photons}

\author{Gubarev F.\,V.}
\email{gubarev@datadvance.net}
\affiliation{DATADVANCE LLC, Russia 117246, Nauchny pr.~17, 15~fl., Moscow}
\affiliation{Quantum Technology Centre, Faculty of Physics, Lomonosov Moscow State University, Moscow, Russian Federation}
\author{Dyakonov I.\,V.}
\author{Saygin M.\,Yu.}
\author{Struchalin G.\,I.}
\author{Straupe S.\,S.}
\author{Kulik S.\,P.}
\affiliation{Quantum Technology Centre, Faculty of Physics, Lomonosov Moscow State University, Moscow, Russian Federation}

\date{\today}

\begin{abstract}
We present a novel semi-analytical methodology to construct optimal linear optical circuits for heralded production of 3-photon GHZ and 2-photon Bell states. We provide a detailed description and analysis of the resulting optical schemes, which deliver success probabilities of 1/54 and 2/27 for dual-rail encoded 3-GHZ and Bell states generation, respectively. Our results improve the known constructive bounds on the success probabilities for 3-GHZ states and are of particular importance for a ballistic quantum computing model \cite{Rudolph2015}, for which these states provide an essential resource.

\end{abstract}

\maketitle
\section{Introduction}

Contemporary quantum computing technology is capable of engineering quantum devices operating with tens of qubits. Various physical platforms are competing in a race to implement quantum algorithms in practice. The linear optical platform is attractive in many ways but suffers from a major hindrance~-- a probabilistic nature of multiqubit gates~\cite{Lutkenhaus1999}. Even though current theoretical proposals explore ways to seamlessly incorporate non-deterministic entangling gates~\cite{Eisert2007}, the currently known linear optical quantum computer (LOQC) architectures still have to consume small entangled states as a resource for successful operation. Current state-of-the-art model for linear-optical quantum computing requires a deterministic source of entangled 3-photon states of the Greenberger-Horne-Zeilinger (GHZ) type \cite{Rudolph2015}. There are approaches to deterministic generation of such states~\cite{Gershoni_Science2016}, however
high-quality deterministic preparation is still out of reach for the current technology. Alternatively, active-multiplexing and heralded entangled state generation circuits provide a solution to the problem at the cost of additional resources. A cornerstone of this approach is the success probability of the entangling gate used, which determines the required volume of supplementary resources.

Probabilistic entangling gates may be divided in two classes~-- postselected and heralded ones. Successful operation of a postselected gate is identified post-factum at the latest stage of an experiment and requires detection of all photons in the circuit. Such gates cannot be concatenated since the input of each gate has to be encoded exclusively in the logical basis~-- a requirement, which is impossible to fulfill due to the unitary behaviour of the circuit \cite{Knill2003}. In other words, the output of any postselected gate with non-unity success probability will
contain unwanted states (often outside the logical basis), which will ruin the operation of the consequent gates. Furthermore, recent work \cite{Adcock2019} provides evidence that postselected entangling gates cannot span the full space of multi-photon entangled states. Last, but not least, the postselected setting demands the photons to pass through the whole circuit which poses extremely stringent requirements on loss in the optical circuit. For the reasons above the use of postselected gates for scalable quantum computing appears to be infeasible. However we must note that in specific small-scale cases interconnection between different degrees of freedom of a single photon can help to overcome the issue \cite{Lanyon2008}.

In turn the heralded gates use some of the input photons to trigger the successful operation event without detecting and thus destroying the photons carrying the logical information. The heralding principle enables a completely new strategy for a linear optical quantum computer architecture: since the successful trigger event exists, the photons carrying the logical information can be measured during the circuit operation and not at the very end of it. The consequence is a drastic increase of the tolerable loss in the optical circuit. It has been shown that a specific large-scale cluster state generation procedure tolerates a few percents of photon loss~\cite{Rudolph2015, Pant2017}. Throughout the rest of the paper we will discuss only the problem of designing heralded entangling gates. 

Let us briefly review the existing results on the optical circuit for heralded entangled state generation. We will focus on the circuits using non-entangled ancillary photons and dual-rail encoded logical qubits. The first example of the heralded CZ gate was reported in the seminal work by Knill et al.~\cite{Knill2001} and had success probability of $1/16$. Later on Knill devised a CZ gate circuit with 2 ancillary single photons~\cite{Knill2002} succeeding with probability $2/27$ and a loose upper bound for any linear optical CZ gate of $3/4$~\cite{Knill2003}. The best result for Bell-state generation is due to Zhang et~al.~\cite{Pan2008} who experimentally demonstrated a circuit with success probability of $3/16$. The GHZ states \cite{GHZ} are substantially harder to generate and few results are known for the general case \cite{Uskov2015, Wehner2019}. To our knowledge, the best result for heralded 3-GHZ generation is reported in~\cite{Varnava2008} and guarantees the success probability of $1/256$ without feedforward and $1/32$ if feedforward is allowed. 

The step-by-step recipe for designing a circuit implementing a particular multiqubit linear optical gate does not exist. A few examples of insights on linear optical gate construction may be found in the literature \cite{Knill2001, Knill2002, Pan2008}. However, for a general problem of finding an optical circuit guaranteeing maximal gate success probability, a more generic approach should be considered. For instance, the linear optical transformation of the input Fock state may be described in terms of a system of polynomial equations~\cite{vanMeter2007}, which has a well-known numerical solution~-- the Buchberger algorithm~-- unfortunately, with an EXPSPACE complexity. 

Better performance may be achieved by formulating the circuit design task as an optimization problem, which fits the unitary transformation of the circuit to a desired quantum gate and minimizes some figure of merit, for example, fidelity of the desired gate and the current gate computed during the procedure \cite{Dowling2009}. This methodology of the quantum gate design is highly sensitive to the details of numerical optimization problem setup and thus requires accurate formulation.

Here we report a detailed analysis of the numerical optimization procedure of finding linear optical circuits for heralded generation of a 3-GHZ state. As a result we present a circuit for dual-rail encoded 3-qubit GHZ state generation with probability of $1/54$ representing a nearly five-fold improvement over the best known result~\cite{Varnava2008}. A part of this circuit may be used to generate two-qubit Bell states with probability of $2/27$. Both circuits do not require any feedforward. 

\section{Problem Setup}

\begin{figure}[t]
\centerline{
\includegraphics[width=0.25\textwidth]{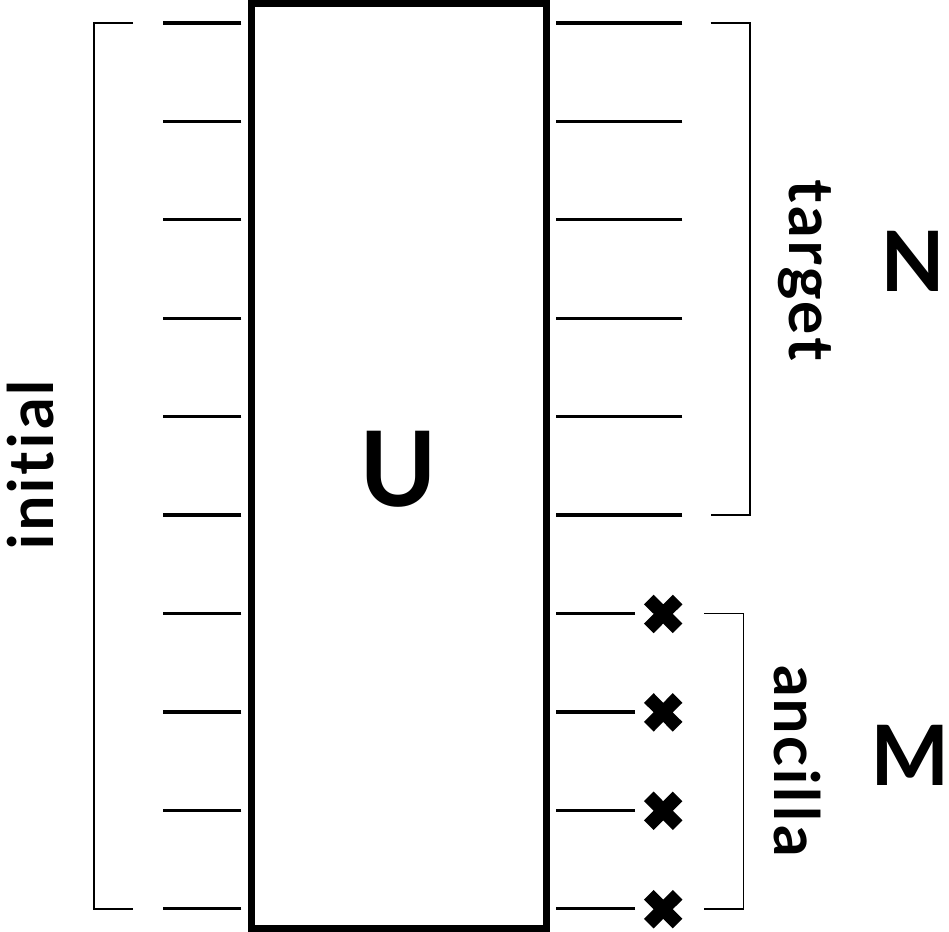}}
\caption{Generic problem setup illustrated for the case $N=N_{ph} = 6$, $M=4$.}
\label{fig::setup}
\end{figure}

We consider a problem of finding a unitary transformation $\mathcal{U}$ of an initial
separable state of $N_{ph}$ photons in $N+M$ modes,
\beq
\label{initial_state}
\bra{\psi_{in}} = \prod_{k=1}^{N_{ph}} \create{a}_{i_k} \,\, \bra{0}^{\otimes[N+M]}\,,
\eeq
such that particular measurement patterns in $M$ ancillary modes
herald the desired $N$-mode target states with maximal success probability
(see Fig.~\ref{fig::setup}). In particular, we will be most interested in maximally entangled 2- and 3-photon target states (Bell-like and GHZ families),
assuming the detectors are capable to distinguish zero, one, and more than one photons, and focus on single-photon ancillary states. The transformation $\mathcal{U}$ of the photonic state in the Fock space corresponds to a unitary transformation $U$ of the annihilation operators, describing an underlying $N+M$ mode interferometer. Matrix elements of $\mathcal{U}$ are related to permanents of the matrix $U$~\cite{Scheel_Arxiv2004}. 

Quantitatively, there are two objects to consider:
probability
$P_a = \sum_m |\ket{ m,a } \mathcal{U} \bra{\psi_{in}}|^2$
of ancillary state $\bra{a}$ detection and corresponding overlaps
$\cM_{t,a} = P_a^{-1} \, |\ket{ t,a} \mathcal{U} \bra{\psi_{in}}|^2$
of the heralded wave function with target vectors,
where $\bra{m,a}$ denotes a normalized Fock-space state $\bra{m_1,\dots m_N,a_1,\dots,a_M}$
with $M_{ph} = \sum_i a_i $ being the number of ancillary photons.
Post-selection (probability $P_a$) of a particular target state means that the $a$-th column
of 
the $\cM$ matrix has a single unit element (remaining entries are zero),
for multiple targets overall success probability is a sum of all appropriate $P_a$.
Therefore, the goal is to find both the optimal set $\mathcal{A}(U)$ of admissible ancillary states
and the corresponding unitary transformation $U$ of optical modes
\begin{gather}
\label{problem1}
\mathcal{A}(U) = \{ \,a\, | \, \exists \, t^*\,: \,\,\cM_{t^*,a}(U) = 1 \,\}\,, \\
U ~=~ \mathrm{arg} \, \max_V\, \sum_{a\in \mathcal{A}(V)} P_a(V)\,. \nonumber
\end{gather}
Note that the solution is not expected to be unique, therefore, it makes sense
to augment the problem with additional performance measure(s).
A natural choice comes from practical considerations: among various equivalent
solutions the ``simplest'' one is preferable, where ``simplicity'' is defined as the minimal number
of non-trivial $U(2)$ factors (optical elements) required to realize a given unitary transformation.
There are various ways to factorize unitary matrices~\cite{Hurwitz1897,Jarlskog2005,Dita1982,Ivanov2008,Reck1994,Clements,Saygin2020,Fldzhyan2020},
here we stick exclusively with the approach of Ref.~\cite{Clements}
\beq
\label{clements}
U(N) \ni U = D \, \cdot \, T^{(n_1,m_1)}_1 \,\, \dots \,\, T^{(n_Q,m_Q)}_Q\,,
\eeq
where $Q = N(N-1)/2$, $D$ is a diagonal matrix of pure phases, and
$T^{(n,m)}$ are $U(2)$ rotations (two-mode ``splitters''):
\begin{gather}
\label{beamsplitter}
T ~=~ \left[ \begin{array}{cc}  e^{i\varphi} \cos\theta & -\sin\theta \\
                                e^{i\varphi} \sin\theta &  \cos\theta \\
\end{array}\right]\,, \\
\theta\in [0,\pi/2]\,, \,\, \varphi \in [-\pi,\pi] \nonumber
\end{gather}
embedded in $(n, m)$ rows/columns.
The transformation $T$ becomes trivial at $\theta = \{0, \pi/2\}$~-- up to a global phase
it reduces either to an identity or a permutation matrix.
Therefore, an additional performance measure to be minimized is
\begin{gather}
S(U) = \sum_i\left\{ (1 - \cos[4 \theta_i]) \,+\, \varepsilon \, (1 - \cos[2 \varphi_i]) \right\} \, +  \nonumber \\
+ \, \delta \sum_i |D_i - 1|^2\,,
\label{problem2}
\end{gather}
where $\varepsilon$ and $\delta$ are small parameters, which gently push the respective phases towards ``trivial'' values (e.g., $\varphi_i = 0, \pm\pi$).

\section{Solution Methodology}

We solve the above described problem using numerical optimization methods,
supplemented with analytic 
post-processing of the results.
Our methodology consists of two main stages:
\begin{enumerate}
\item A particular approximate solution of (\ref{problem1}) is obtained
      using numerical methods (see below for details);
\item Once the candidate ancillary indices $\mathcal{A}$ are established, we numerically solve (\ref{problem2})
      supplemented with appropriately lower-bounded ancilla probabilities $P_a$, $a \in \mathcal{A}$ and
      corresponding requirements on the overlap matrix elements.
\end{enumerate}
These steps are repeated multiple times to assure a global search of an optimal solution, the best results obtained are collected for further processing.

A silent feature of (\ref{problem1}) is that it does not admit a direct
formulation as a constrained optimization problem because neither the relevant
set of ancillary indices $A$ nor the target states with unit overlaps are known {\it a priori}.
Theoretically, one could try to introduce additional discrete variables, however,
available methods to solve the resulting non-linear constrained mixed-integer
task are rather inefficient and are likely to reduce to exhaustive enumeration.
Experience revealed that the most efficient approach is to consider, following Ref.~\cite{Stanisic},
\beq
\label{problem1a}
U ~=~ \mathrm{arg} \, \max_U\, \sum_{t,a} P_a \,\, \cM^p_{t,a}\,,
\eeq
where the summation is done over all targets and ancillas and $p$ is some positive power,
which ensures sufficient suppression of small matrix elements $\cM_{t,a}$ (in practice, we used $p=3,4,5$).
The resulting formulation has no explicit constraints and can be solved efficiently.
However, optimal solutions of the original problem (\ref{problem1}) generically become
only local optima of (\ref{problem1a}), therefore, all extremal points of the latter
are to be considered.
Fortunately, this limitation is not very relevant in practice, since the most powerful
gradient-based local optimization methods, which we use, find only local optima anyway.
In more details, the considered objective function is smoothly differentiable almost
everywhere and its derivatives are known analytically.
It follows then that (locally) optimal points are to be found most efficiently
with second-order gradient-based algorithms of (quasi)-Newton family.
Specifically, we utilized a particular modern numerical realization of dumped BFGS method
stabilized with Wolf-like line search rules,
which constitutes an inherent part of the {\it pSeven Core} algorithmic package
(see \cite{datadvance} for more details).
Although other suitable optimization techniques are applicable as well,
their performance in the present context is expected to be much worse.

Another cornerstone of the proposed methodology is a proper parameterization of the unitary group.
In this study an open neighborhood of an arbitrary $U_0 \in U(N+M)$ is parameterized via Cayley transform
\beq
\label{cayley}
U = U_0 \cdot \frac{i - H}{i + H}\,,
\eeq
where $H$ is an $(N+M) \times (N+M)$ Hermitian matrix with $(N+M)^2$ unconstrained real parameters.
Representation \eqref{cayley} is well known in matrix analysis~\cite{golub}
and was proven to be efficient in various applications (see, e.g., Refs.~\cite{cayleyRef1,cayleyRef2}).
In our case, for an arbitrary constant unitary $U_0$ the real parameters of $H$ define the design space for the problem \eqref{problem1a}, exploration of which is to be started at $H = 0$ with local optimization methods. To ensure globalized search we considered tens of thousands Haar-random initial $U_0$, each of which were then locally optimized using the parametrization \eqref{cayley}.
Note that an appropriate solution to \eqref{problem1a} is not always established,
there is a large number of improper stationary points, $\cM_{t,a} \ne 0, 1$,
which do not admit identification of an optimal set $\mathcal{A}(U)$ and thus are to be rejected.

A suitable solution of (\ref{problem1a}) establishes both the subset of ancillary indices $\mathcal{A}$ and
the set of overlaps $\cM_{t,a}$ to be kept at unit value during the second stage.
Therefore, next we consider the problem (\ref{problem2}) supplemented with additional constraints
\beq
\label{problem2a}
P_a \ge P_a^*\,, \qquad \cM_{t^*,a} = 1\,, \qquad a \in \mathcal{A}\,,
\eeq
where $P_a^*$ denotes the ancilla probabilities obtained at the first stage.
In turn, the quality of the second stage solution is given by the minimal number
of non-trivial optical elements within the decomposition (\ref{clements}).
Therefore, at the second stage one has to solve a constrained single-objective optimization task \eqref{problem2},\eqref{problem2a}, solution of which was obtained with a
sequential quadratically constrained quadratic programming (SQCQP) algorithm provided
by {\it pSeven Core}, \cite{datadvance}.
The above two-step procedure is repeated several thousand times with different random starting
points $U_0$ and different powers $p$. The selected set of best $U$'s is leaved for further analytic treatment,
to which we turn next.

\section{3-GHZ States Generation}

\begin{figure}[t]
\centerline{
\includegraphics[width=0.5\textwidth]{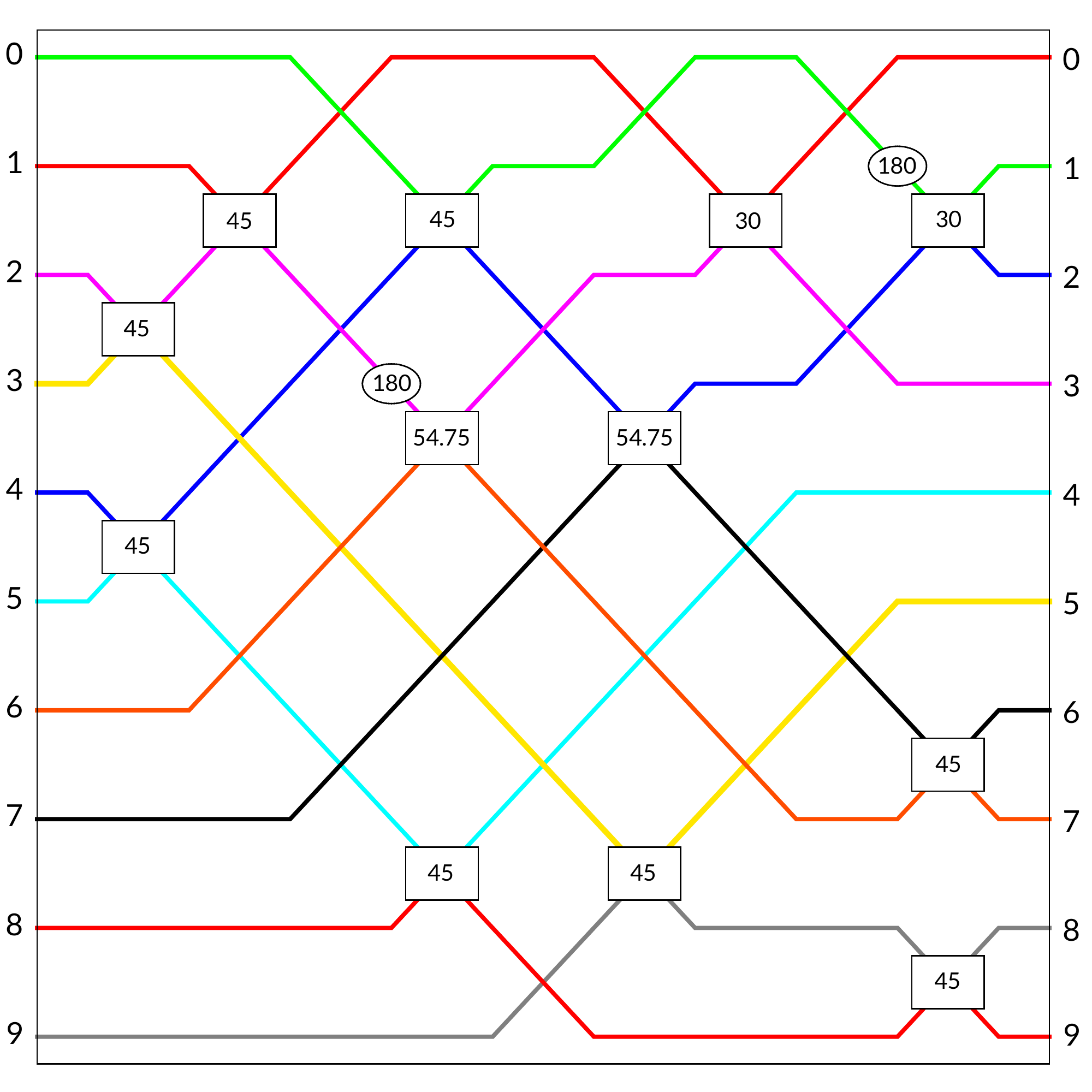}}
\caption{The simplest scheme identified (12 two-mode elements), the ports assignment agrees with Fig.~\ref{fig::setup} ( the input state $\bra{1}^{\otimes 6}\bra{0}^{\otimes 4}$ is on the left).
Ellipses represent single mode phase shifters, boxes stand for splitters (\ref{beamsplitter}) with $\varphi=0$.
All angles are in degrees rounded to the second digit when appropriate, in particular, $54.75^\circ = \mathrm{acos}\frac{1}{\sqrt{3}}$.}
\label{fig::scheme1}
\end{figure}

In this Section we consider the problem of optimal three particle GHZ-states generation using six unentangled photons and four ancillary modes, which corresponds to $N = N_{ph} = 6$, $M = 4$ in the setup in Fig.~\ref{fig::setup}.
Without loss of generality the input state is taken to be
$\bra{\psi_{in}} = \bra{1}^{\otimes 6}\, \bra{0}^{\otimes 4}$.
The target vectors are $\pm$ superpositions of states with 3 photons in 6 modes and all mode occupation numbers being zero or one, e.g.
$\bra{t_k} \propto \bra{100110} \pm \bra{011001}$
and all unique particle number permutations thereof (the second component is a binary complement of the first). Such states correspond to 3-qubit GHZ states in appropriately chosen dual-rail encodings. Admissible measurement patterns include states with 3 photons in 4 ancillary modes, where each mode contains zero or one photon (thus, there are only four legitimate heralding patterns).

Extensive numerical experiments revealed that solutions of (\ref{problem1a})
with the same quality appear quite often, e.g., a Haar-uniform distribution of initial
points $U_0$ results in the same quality transformations in $\sim20\%$ of cases.
Meanwhile, the typical distance $\rho(U^*, U_0) = 1 - {\mathrm Tr}[U^* U^\dagger_0]/(N+M)$ 
between $U_0$ and the reached optimal element $U^*$ is of the order 1/2
(more precisely, it is $\langle\rho\rangle = 0.45(15)$ under a Gaussian approximation to the attained statistics).
A distinguishing feature of all numerically identified unitary matrices is that
two and only two ancillary states have appropriate overlaps with the selected targets.
Specific ancilla indices as well as heralded states might change, however, they always
come in pairs. Extension or reduction of the number of target GHZ states does not change this property,
the corresponding success probabilities remain $P_a = 0.00925926(1) \approx 1/108$ per each
successful measurement.

Candidates tuning via  \eqref{problem2}, \eqref{problem2a} indicated that the transformation complexity varies greatly, the required minimal number of two-mode elements might be as large as $\sim 30$.
However, we attribute this to inherent multimodality of the considered formulation, because of which only locally optimal designs are often identified. Globalization is achieved as usual via selection of the simplest unitaries demonstrating the same performance from those collected in all conducted runs.
It turns out that the minimal attainable number of elementary splitters is $12$, at least we never encountered a better solution. 

Finally, we collected a few dozens of best optical transformations with the number of splitters equal to $12$ and $13$ (which is to be compared with the generic case of $10 (10 - 1)/2=45$ elements).
It turned out that all of them are just the repetitions
(up to permutations of ports and rearrangement of phase shifters)
of the same scheme, presented in Fig.~\ref{fig::scheme1}.
Note that each box in the figure represents a two-mode transformation (\ref{beamsplitter}) taken at $\varphi=0$, ellipses denote single-mode phase shifts (actually, sign flips). The scheme of Fig.~\ref{fig::scheme1} is not a direct result of a numerical experiment, it was obtained with extensive analytic post-processing (phase shifters reduction and removal of unnecessary optical path crossings) and guessing of involved algebraic numbers. Nevertheless, the numeric treatment was invaluable in its determination.

\begin{figure}[t]
\centerline{
\includegraphics[width=0.5\textwidth]{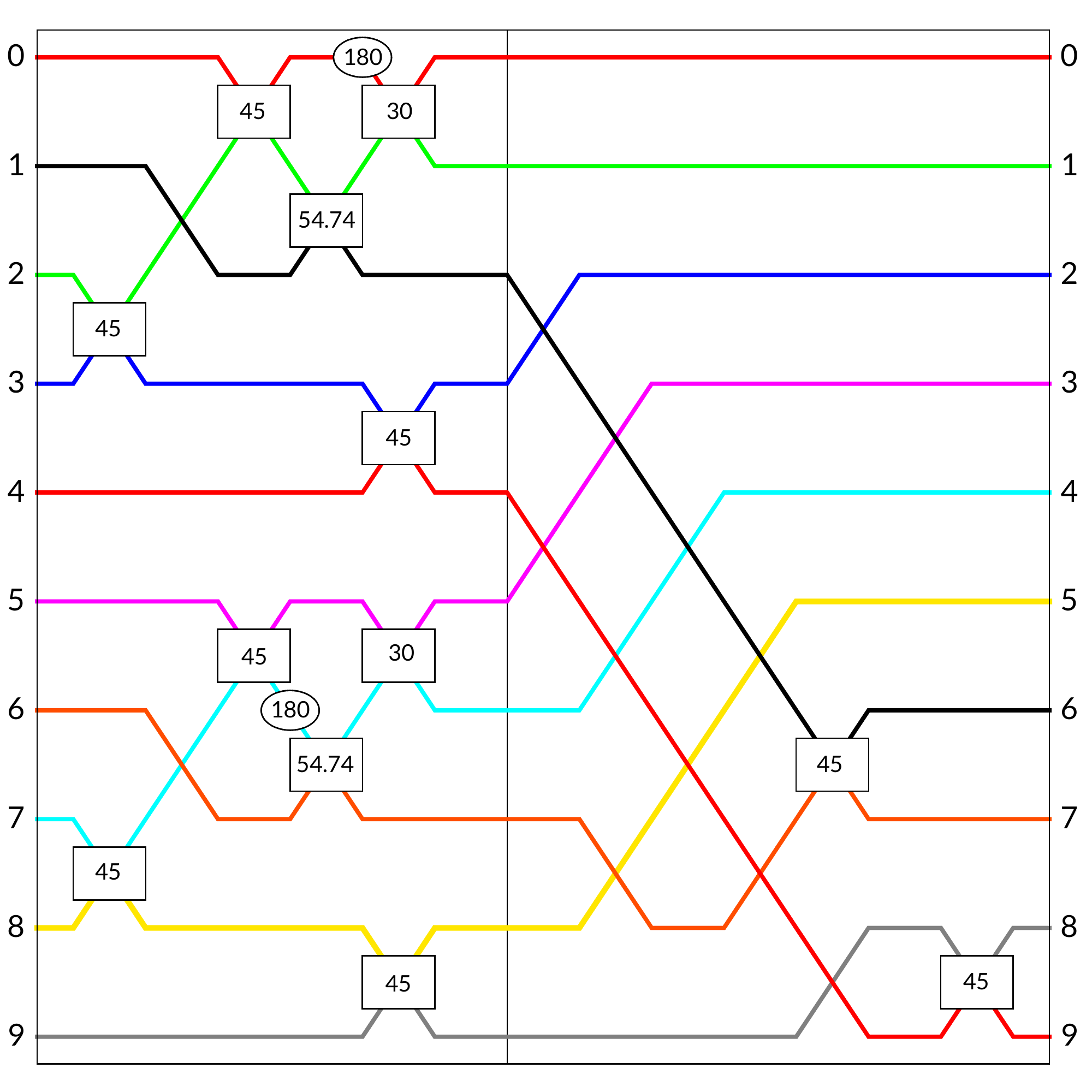}}
\caption{Optical scheme decomposition, obtained via relabeling of the input (left-hand side) ports
so that the appropriate initial state is $\bra{1011010110}$. Output ports are ordered in accordance with
$10 = 6 + 4$ convention, the heralded state is in the first 6 modes. The vertical line separates the disjoint structure from the two-modes mixing elements and the output port permutations.}
\label{fig::parts-combined}
\end{figure}

In fact, the established optical scheme is almost disjoint and consists of two nearly symmetric arms interconnected via a couple of $45^\circ$ splitters at the output. The corresponding decomposition, equivalent to Fig.~\ref{fig::scheme1} up to the input
ports permutation, is shown in Fig.~\ref{fig::parts-combined}, where we kept a conventional 6+4 ordering of the output ports at the expense of a perhaps redundant number of optical path crossings.
Till the end of the current Section we will concentrate on this representation, although it seems not to be the most illuminating: we argue below that further scheme surgery delivers more comprehensive insights and reveals its connections with some known protocols.

An analytic form of the established transformation is easy to derive.
The unitaries, corresponding to the top $U_T$ and bottom $U_B$ disjoint parts (to the right of the vertical line in Fig.~\ref{fig::parts-combined}) are almost the same:
\beq
\label{parts}
U_{T,B} = \left[
\begin{array}{rrrrr}
\myrule         \mp\sqrt{\frac{2}{3}} & \frac{1}{\sqrt{6}}  & \frac{\pm 1}{2\sqrt{3}} & \frac{\mp 1}{2\sqrt{3}}  &    \\
\myrule                             & \frac{-1}{\sqrt{2}} & \frac{\pm 1}{2}         & \frac{\mp 1}{2}          &    \\
\myrule          \frac{\pm 1}{\sqrt{3}} & \frac{1}{\sqrt{3}}  & \frac{\pm 1}{\sqrt{6}}  & \frac{\mp 1}{\sqrt{6}}   &    \\
\myrule                             &                     & \frac{1}{2}         & \frac{1}{2}           &  \frac{-1}{\sqrt{2}} \\
\myrule                             &                     & \frac{1}{2}         & \frac{1}{2}           &  \frac{1}{\sqrt{2}}
\end{array}
\right]\,,
\eeq
where only non-zero matrix elements are shown.
In accordance with Fig.~\ref{fig::parts-combined}, the  total transformation matrix is obtained 
from the block diagonal $\mathrm{diag}[U_T,U_B]$ matrix via a permutation of rows (output ports), left application of two $45^\circ$ splitters (rows mixing) and relabeling of modes to arrive to the $10 = 6 + 4$ convention. The GHZ state is heralded in modes $(0-5)$ if and only if single photons are detected
in both $(6,7)$ and in either one of $(8,9)$ ports (the remaining one is to be found in a vacuum state).
Each event happens with probability $P_a = 1/108$, so that the overall success rate equals to $P_\text{success} = 1/54$.

To justify the above assertion let us note that the transformed state is determined by the polynomial
$(27 \cdot 2^5)^{-1}\, Q(\create{a}_0 \dots \create{a}_9)$ in creation operators:
\begin{gather}
Q = [A^+_{6,7} - 2 \create{a}_0] \cdot [A^-_{6,7} + 2 \create{a}_3] \cdot \\
\cdot [A^+_{6,7} - \sqrt{\frac{3}{2}} A^-_{8,9} + C^{(1)} ] \cdot [A^-_{6,7} + \sqrt{\frac{3}{2}} A^+_{8,9} + C^{(3)} ] \cdot \nonumber \\
\cdot [A^+_{6,7} + \sqrt{\frac{3}{2}} A^-_{8,9} + C^{(2)} ] \cdot [A^-_{6,7} - \sqrt{\frac{3}{2}} A^+_{8,9} + C^{(4)} ] \,, \nonumber \\
C^{(1,2)} =   \create{a}_0 + \sqrt{3} (\create{a}_1 \pm \create{a}_2)\,, \nonumber \\
C^{(3,4)} = - \create{a}_3 - \sqrt{3} (\create{a}_4 \mp \create{a}_5)\,, \nonumber
\end{gather}
where  $A^\pm_{i,j} = \create{a}_i \pm \create{a}_j$.
The required measurement patterns correspond to the products
$\create{a}_6 \create{a}_7 \create{a}_8$ and
$\create{a}_6 \create{a}_7 \create{a}_9$,
which are entirely contained in the terms proportional to
$(\,A^\pm_{6,7}\,)^2 \, A^\pm_{8,9}$ (all sign combinations).
The structure of the above expression reveals that the
monomials $(\,A^+_{6,7}\,)^2 \, A^-_{8,9}$ and $(\,A^-_{6,7}\,)^2 \, A^+_{8,9}$
enter with zero coefficients.
The coefficients of the remaining same-sign monomials are given by:
\begin{gather}
(A^+_{6,7})^2 \, A^+_{8,9} = 2 \create{a}_6 \create{a}_7 (\create{a}_8 + \create{a}_9)\,\,: \\
2 \sqrt{\frac{3}{2}} \left(C^{(4)} - C^{(3)} \right)
  \left( \create{a}_3\left[C^{(1)} + C^{(2)}\right] - 2 \create{a}_0 \create{a}_1 \right) \nonumber \\
= \frac{-1}{\sqrt{2}} (2\sqrt{3})^3\, \create{a}_1 \create{a}_3 \create{a}_5\,, \nonumber
\end{gather}
\begin{gather}
(A^-_{6,7})^2 \, A^-_{8,9} = 2 \create{a}_6 \create{a}_7 (\create{a}_9 - \create{a}_8)\,\,: \\
2 \sqrt{\frac{3}{2}} \left(C^{(2)} - C^{(1)} \right)
  \left( \create{a}_0\left[C^{(3)} + C^{(4)} \right] + 2 \create{a}_0 \create{a}_3 \right) \nonumber \\
= \frac{1}{\sqrt{2}} (2\sqrt{3})^3\, \create{a}_0 \create{a}_2 \create{a}_4 \nonumber \,.
\end{gather}
Therefore, the relevant terms in the transformed state are readily obtained
\begin{align}
\create{a}_6 \create{a}_7 \create{a}_8 \, :  \quad & \frac{-1}{6\sqrt{3}} \cdot \frac{(\create{a}_0 \create{a}_2 \create{a}_4 + \create{a}_1 \create{a}_3 \create{a}_5 )}{\sqrt{2}}\,, \\
\create{a}_6 \create{a}_7 \create{a}_9 \, :  \quad & \frac{1}{6\sqrt{3}} \cdot \frac{(\create{a}_0 \create{a}_2 \create{a}_4 - \create{a}_1 \create{a}_3 \create{a}_5 )}{\sqrt{2}}\,,
\end{align}
confirming our assertion.

\section{Scheme Analysis and Bell States Generation}

The above considerations were somewhat formal and only show that the established scheme operates properly, producing 3-GHZ states with success probability of $1/54$. In this Section we perform a more detailed analysis and generalize the scheme to the case of maximally entangled two-photon Bell states generation. It turns out that the resulting transformation shares some similarities with known protocols. Specifically, we will demonstrate how a well-known result of $2/27$ probability for Bell-state generation \cite{carolan} is reproduced using the building blocks identified in the 3-GHZ generation circuit.

\begin{figure}[t]
\centerline{
\includegraphics[width=0.5\textwidth]{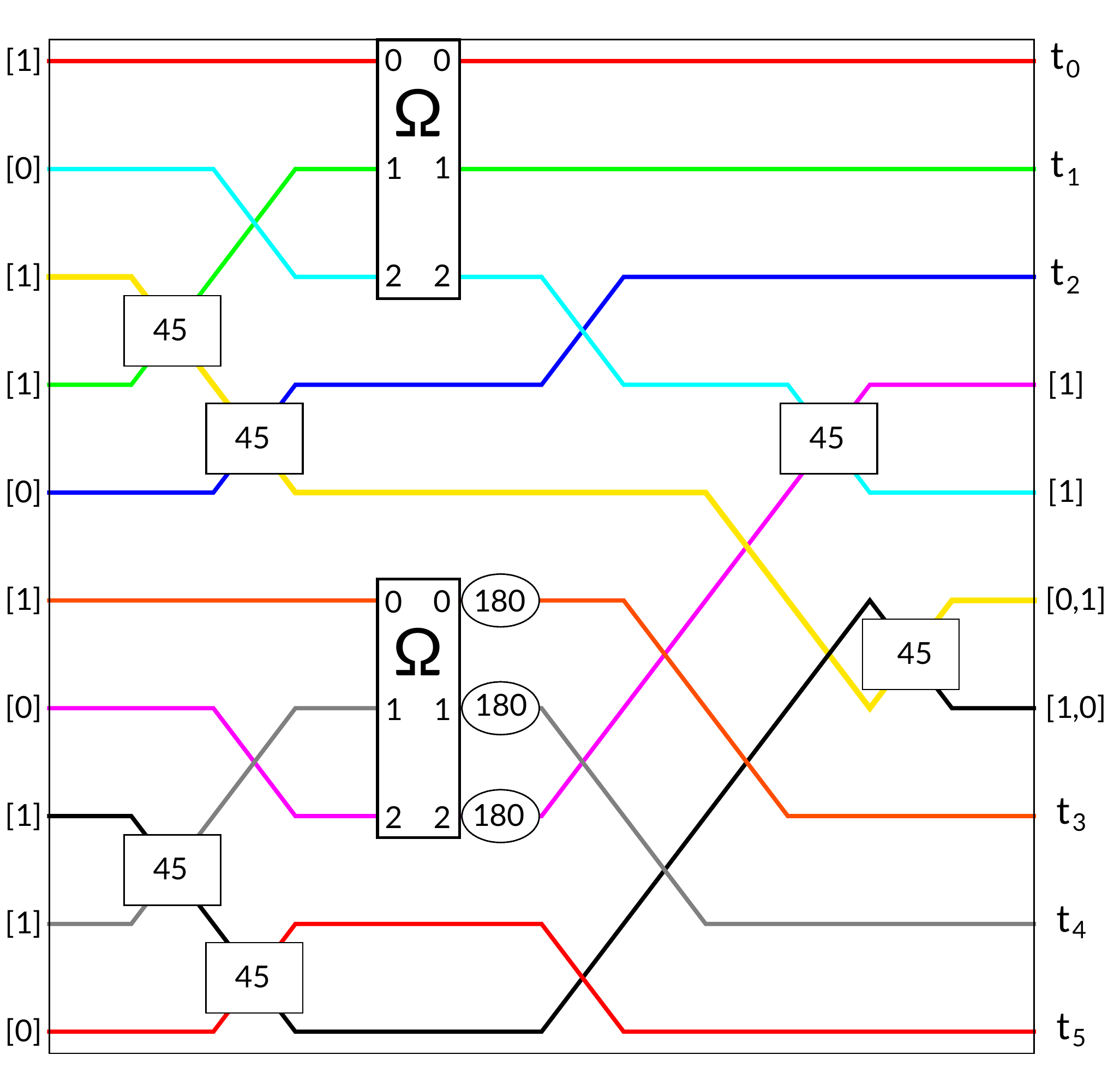}}
\caption{Representation of the 3-GHZ states generation scheme equivalent to Fig.~\ref{fig::parts-combined}. The output ports are $t_0\dots t_5$, the inputs are marked with appropriate photon numbers. The ancillary modes measurement patterns (on the left) are labelled with heralding photon counts.
$45^\circ$ elements are the matrices \eqref{beamsplitter} taken at $\varphi = 0$, ellipses represent
$\pi$-phase shifts, $\Omega$ blocks are detailed in Fig.~\ref{fig::omega}.}
\label{fig::surgery1}
\end{figure}

\begin{figure}[t]
\centerline{
\includegraphics[width=0.5\textwidth]{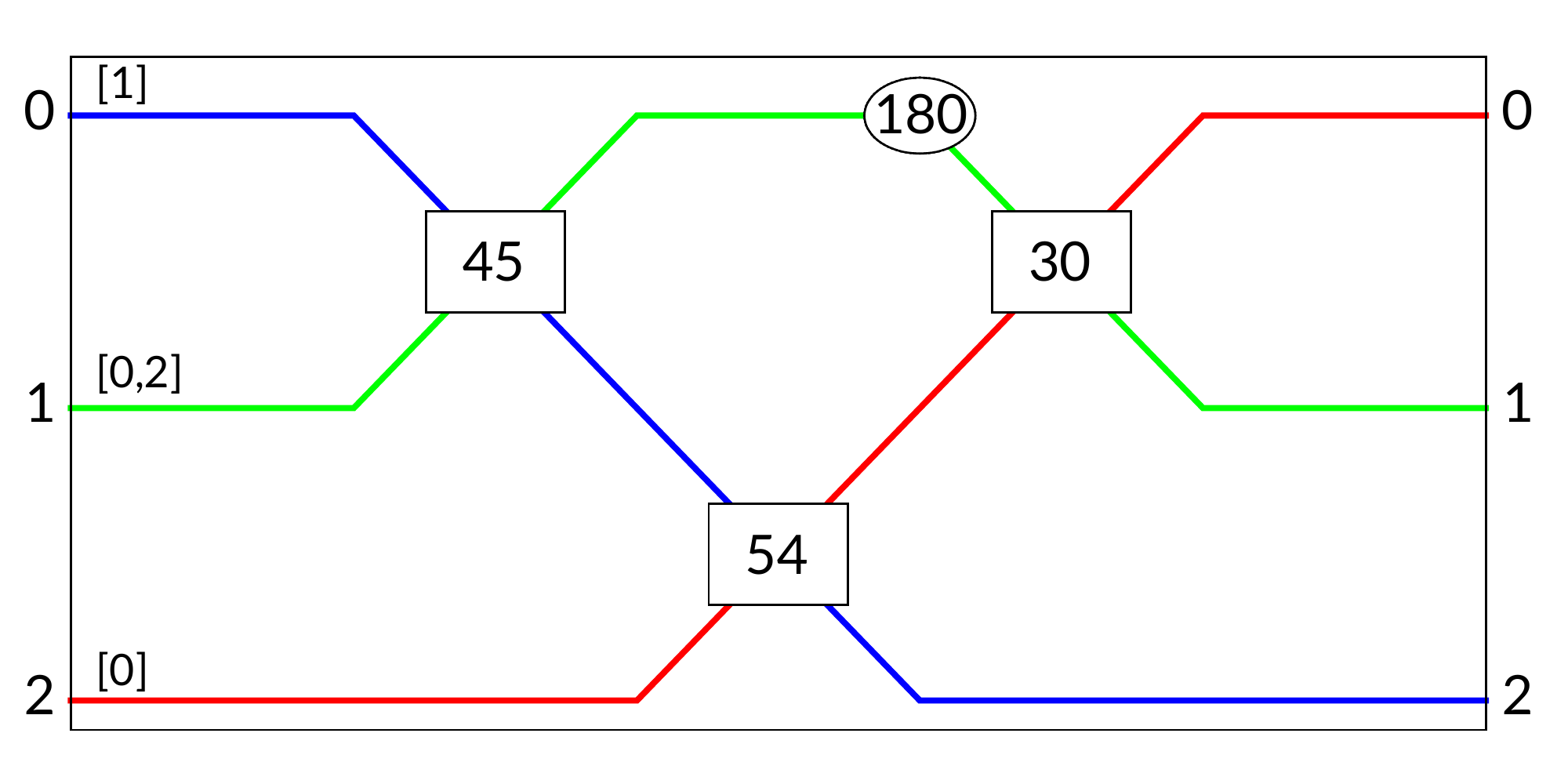}}
\caption{$\Omega$ block of considered optical circuit, the
appropriate number of photons is indicated in square brackets for each input port.}
\label{fig::omega}
\end{figure}

Fig.~\ref{fig::surgery1} represents a useful dissection of the optical scheme described above. Note that the heralded state (3-GHZ in this case) is to be found in the target ports $t_0 \dots t_5$, the remaining ancillary modes are marked with the appropriate measured photon numbers, the input occupation numbers are indicated explicitly (all other conventions are the same as before).
$\Omega$ blocks to be discussed shortly are defined in Fig.~\ref{fig::omega}, from which it follows that the scheme of Fig.~\ref{fig::surgery1} is identical to what we considered previously.

Despite its rather complex look, Fig.~\ref{fig::surgery1} admits a straightforward interpretation. Indeed, the four left-most $45^\circ$ splitters are designated to prepare an appropriate input state to be further processed in the $\Omega$ blocks, while the two analogous right-most devices are dedicated
to heralding measurements. One can observe that the input ports $\Omega.1$ can receive zero or two photons only due to the Hong-Ou-Mandel interference at the preceding beamsplitters~\cite{Mandel}. For the same reasons and because of the indicated heralding measurements only the coherent superpositions of zero and two photons at the output ports $\Omega.2$ are relevant. Therefore, the scheme functions via a coordinated operation of two $\Omega$ blocks, while the complicated combination of $45^\circ$ splitters is somewhat auxiliary and, in fact, is well known. It was proposed in Ref.~\cite{zou-pahlke} for multi-photon GHZ states production
and provides the success probability of $1/64$ in the three-photon case.

\begin{figure}[t]
\centerline{
\includegraphics[width=0.5\textwidth]{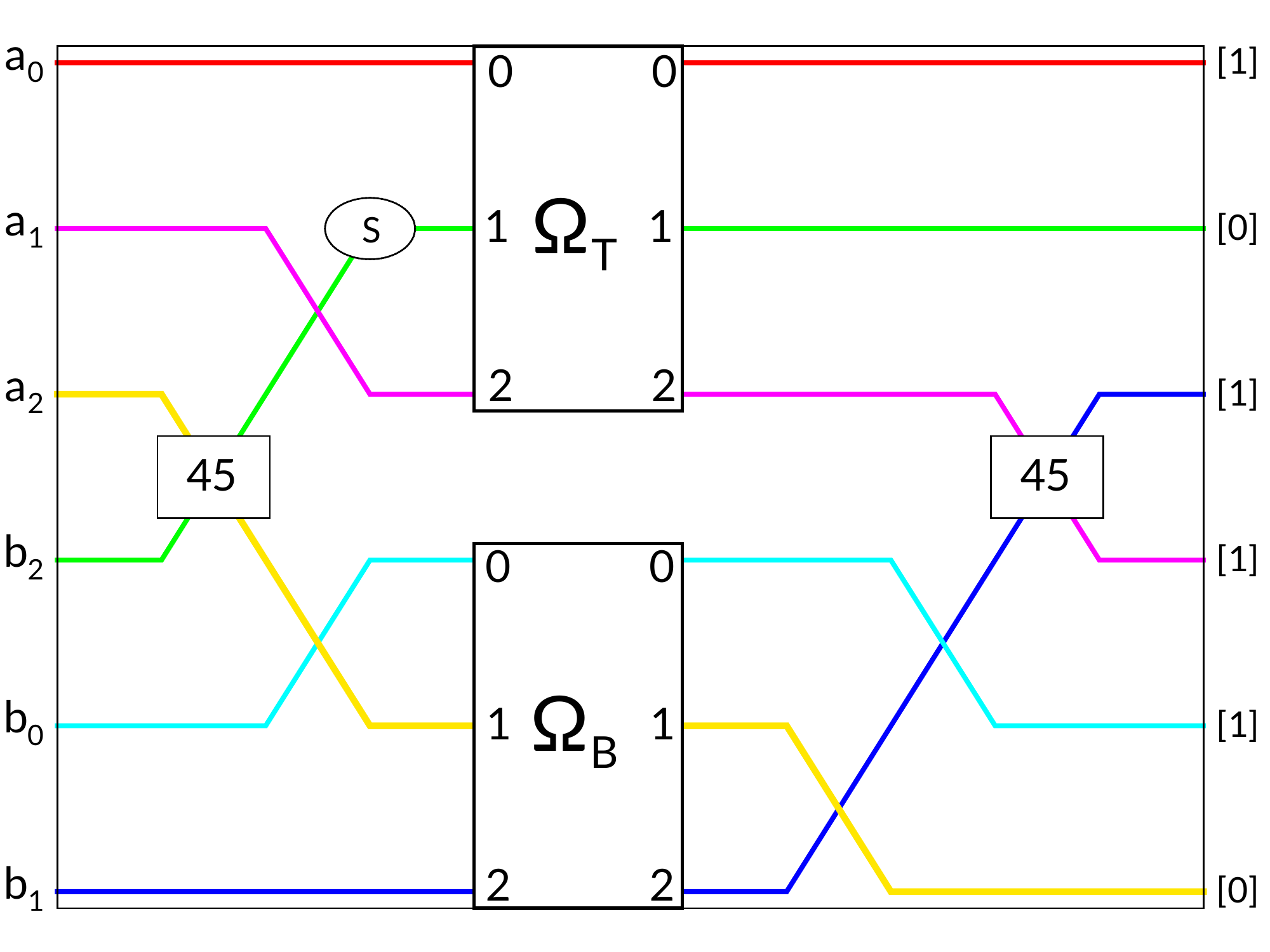}}
\caption{Scheme to generate maximally entangled Bell states of two photons
(ports $(a_0,a_1)$ and $(b_0,b_1)$) using $\Omega$ blocks from Fig.~\ref{fig::omega}.
The ancillary modes are $a_2$, $b_2$, input photon numbers are indicated in square brackets.}
\label{fig::bell}
\end{figure}

Therefore, it makes sense to consider a simplified version of the scheme, Fig.~\ref{fig::bell}, which is aimed to produce Bell states. The only new element is an optional $\pi/2$-phase shift, marked as an $s$-circle ($s=\pm 1$ or simply $s=\,\pm$), which implements a sign flip of a two-photon component amplitude, $\bra{2} \to s \bra{2}$. Qualitatively the circuit operates in the same way as before, however, its quantitative characteristics deserve a separate discussion.
Qubits $a$ and $b$ are identified with the output mode pairs $(a_0,a_1)$, $(b_0,b_1)$ and we use a 
set of Bell states associated with this dual-rail encoding:
$\bra{\phi^\pm} \propto (\create{a}_0 \create{b}_0 \pm \create{a}_1 \create{b}_1)\bra{0}$,
$\bra{\psi^\pm} \propto (\create{a}_0 \create{b}_1 \pm \create{a}_1 \create{b}_0)\bra{0}$.
Below we will need to analyze a few different $\Omega$-like blocks, therefore, in accordance with expected input states and heralding measurements, it is worth to consider a generic transformation of $\create{\omega}_0$, $(\create{\omega}_1)^2$:
\begin{gather}
\label{mode-transform-0}
\create{\omega}_0 \, \to \, \alpha \, \create{\omega}_2 + \beta_\omega \\
\frac{1}{2} (\create{\omega}_1)^2 \, \to \, A (\create{\omega}_2)^2 + 2 B_\omega \create{\omega}_2 + C_\omega\,, \nonumber
\end{gather}
where $\alpha$, $A$ are block-specific numbers and $B_\omega$, $\beta_\omega$ [$C_\omega$]
are linear [quadratic] in $\create{\omega}_{0,1}$ creation operators, naturally assigned to respective modes.
The transformation rule for $\create{\omega}_0 (\create{\omega}_1)^2/2$ reads:
\begin{multline}
\label{mode-transform-1}
\frac{1}{2} \create{\omega}_0 (\create{\omega}_1)^2 \, \to \, (\create{\omega}_2)^0 \cdot (\dots) + (\create{\omega}_2)^1 \cdot (D_\omega) + \\
  + (\create{\omega}_2)^2 \cdot (A \beta_\omega + 2 B_\omega \alpha) + (\create{\omega}_2)^3 \cdot (\dots)\,,
\end{multline}
where $D_\omega = \alpha C_\omega + 2 B_\omega \beta_\omega$ and dots denote unimportant terms.
Indeed, in the proposed scheme of Fig.~\ref{fig::bell}
heralding measurements require two photons in modes $a_2$, $b_2$
and select only the terms quadratic in $\create{\omega}_{T,2}$, $\create{\omega}_{B,2}$ operators.
After the action of the very first $45^\circ$ splitter the state is proportional to
$\create{\omega}_{T,0} \create{\omega}_{B,0} [ (\create{\omega}_{T,1})^2 - s (\create{\omega}_{B,1})^2 ]$,
from which it follows that quadratic in $\create{\omega}_{T,2}$, $\create{\omega}_{B,2}$ expressions cannot arise from $\sim (\create{\omega}_2)^0$, $\sim (\create{\omega}_2)^3$ terms in Eq.~(\ref{mode-transform-1}).
Finally, accounting for the left-most $45^\circ$ splitter
and naturally identifying $\create{\omega}_{T,k} = \create{a}_k$, $\create{\omega}_{B,k} = \create{b}_k$, $k=\{0,1\}$
one obtains the following expressions in front of the relevant operators:
\beq
\label{generic-expansion}
\create{a}_2 \create{b}_2\,:\,\, -[ (1+s)A \, \beta_a\beta_b + 2\alpha (B_b\beta_a + s B_a\beta_b)]\,,
\eeq
\begin{align}
\frac{(\create{a}_2)^2}{\sqrt{2}}, \frac{(\create{b}_2)^2}{\sqrt{2}} & \,:\,\,
   \frac{1}{\sqrt{2}}[ (1-s)A \beta_a\beta_b + \\
  & + 2\alpha (B_b\beta_a - s B_a\beta_b) \mp \alpha (D_b - s D_a)] \nonumber
\end{align}

\noindent
For the $\Omega$ block illustrated in Fig.~\ref{fig::omega} the
respective unitary transformation matrix is
\beq
\label{omega-U}
U_\Omega ~=~ \left[ \begin{array}{ccc}
                       -\sqrt{2/3} & 1/\sqrt{6} & 1/\sqrt{6} \\
                        0          & 1/\sqrt{2} & -1/\sqrt{2} \\
                        1/\sqrt{3} & 1/\sqrt{3} & 1/\sqrt{3}
 \end{array} \right]\,,
\eeq
from which it follows that
\begin{gather}
\alpha = 1/\sqrt{3}\,, \quad A = 1/6\,, \quad \beta_a = -\sqrt{2/3} \, \create{a}_0\,,\\
B_a = \create{\tilde{a}}_0/3\sqrt{2}\,, \quad C_a = (\create{\tilde{a}}_1)^2/3\,, \quad
D_a = \create{\tilde{a}}_0 \create{\tilde{a}}_1/3\,, \nonumber
\end{gather}
where operators $\create{\tilde{a}}_{0,1}$ are related to $\create{a}_{0,1}$ via a $60^\circ$ rotation,
$\create{\tilde{a}}_{0,1} = \create{a}_{0,1}/2 \pm \create{a}_{1,0}\sqrt{3}/2$.
Using \eqref{generic-expansion} one derives
\beq
\label{omega-expansion-1}
\create{a}_2 \create{b}_2\,:\,\, \frac{\sqrt{2}}{3\sqrt{3}} \cdot
   \frac{\create{a}_0 \create{b}_1 + s \create{a}_1 \create{b}_0}{\sqrt{2}} ~=~
\frac{\sqrt{2}}{3\sqrt{3}} \cdot \bra{\psi^s}\,,
\eeq
which implies that the measurement of the $\bra{1_{a_2} 1_{b_2}}$ ancillary state heralds the maximally entangled state of $a$ and $b$ qubits for either choice of sign $s$ with success probability of $2/27$.

\begin{figure}[t]
\centerline{
\includegraphics[width=0.5\textwidth]{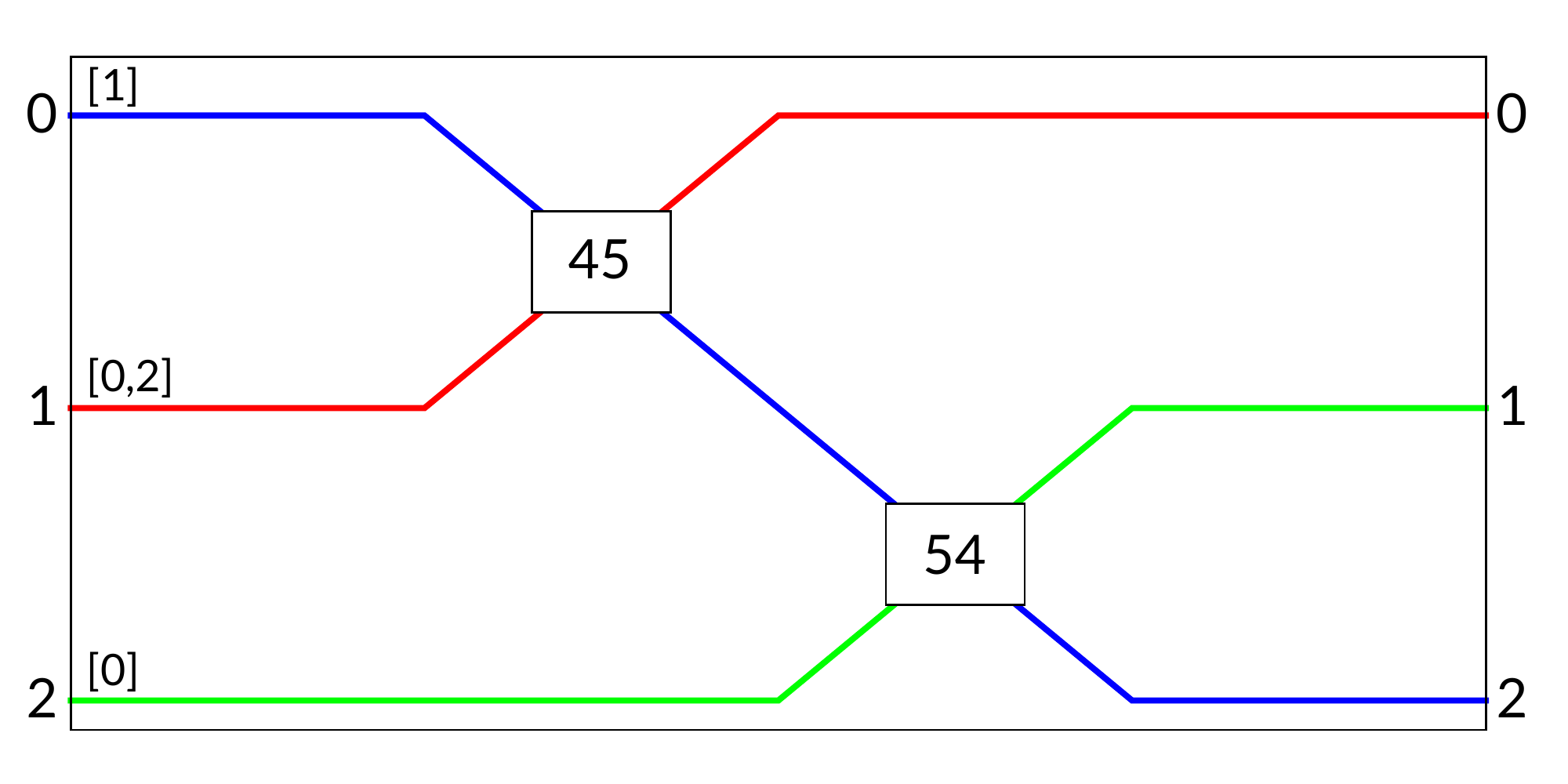}}
\caption{Modified block $\Omega'$, which corresponds to \eqref{omega-modified-U}.
It leads to \eqref{omega-modified-expansion-1}, when used within the circuit of Fig.~\ref{fig::bell}.}
\label{fig::omega-modified}
\end{figure}

Finally, let us consider the significance of the right-most $30^\circ$ splitter in the $\Omega$ block discussed above. Upon removal of this optical element the scheme in Fig.~\ref{fig::bell} becomes similar to the one,
presented in Ref.~\cite{carolan}, which, as claimed, operates with a $2/27$ success rate~\footnote{
Unfortunately, it appears impossible to reproduce the details of the cited circuit:
 the paper itself provides insufficient details, while the unitary matrix given in the supplementary materials is misprinted and does not correspond to the required transformation.
}. 
Note however that this similarity is somewhat formal:  in the scheme of Ref.~\cite{carolan} the coherent superposition of $\bra{02}$ and $\bra{20}$ states is fed to the $54^\circ$ splitters (the input modes of $\Omega.2$ in Fig.~\ref{fig::omega} and \ref{fig::bell}), while our scheme requires it to be in $\Omega.1$.
In either case, the modified $\Omega'$ block of Fig.~\ref{fig::omega-modified} corresponds to
\beq
\label{omega-modified-U}
U_{\Omega'} ~=~ \left[ \begin{array}{ccc}
                       1/\sqrt{2} & -1/\sqrt{2} & 0 \\
                       1/\sqrt{6} &  1/\sqrt{6} & -\sqrt{2/3} \\
                       1/\sqrt{3} & 1/\sqrt{3} & 1/\sqrt{3}
 \end{array} \right]\,,
\eeq
from which one obtains the coefficients in the transformed wave function relevant for heralding measurements:
\begin{gather}
\label{omega-modified-expansion-1}
\create{a}_2 \create{b}_2\,:\,\, \frac{-\sqrt{2}}{3\sqrt{3}}\cdot
\left\{
{\scriptstyle
\begin{array}{cc}
\frac{1}{2} \bra{\psi^+} - \frac{\sqrt{3}}{2} \bra{\phi^-}\,, & s=+ \\ \myrule
            \bra{\psi^-}\,,                                   & s=-
\end{array}
}
\right.
\\
\frac{(\create{a}_2)^2}{\sqrt{2}}  \,:\,\, \frac{\sqrt{2}}{3\sqrt{3}} \cdot
\left\{
{\scriptstyle
\begin{array}{cc}
(\bra{\tilde{\psi}^-} + \bra{\tilde{\chi}^-})/\sqrt{2}\,, & s=+ \myrule \\ \myrule
\frac{\frac{\sqrt{3}}{2}\bra{\tilde{\phi}^-} + \frac{1}{2} \bra{\tilde{\psi}^+} - \bra{\tilde{\chi}^+}}{\sqrt{2}}\,, & s=-
\end{array}
}
\right. \nonumber
\\
\frac{(\create{b}_2)^2}{\sqrt{2}} \,:\,\, \frac{\sqrt{2}}{3\sqrt{3}} \cdot
\left\{
{\scriptstyle
\begin{array}{cc}
(\bra{\tilde{\psi}^-} - \bra{\tilde{\chi}^-}) /\sqrt{2}\,, & s=+ \myrule \\ \myrule
\frac{\frac{\sqrt{3}}{2} \bra{\tilde{\phi}^-} + \frac{1}{2} \bra{\tilde{\psi}^+} + \bra{\tilde{\chi}^+}}{\sqrt{2}}\,, & s=-
\end{array}
}
\right., \nonumber
\end{gather}
where $\bra{\chi^\pm} \propto (\create{a}_0 \create{a}_1 \pm \create{b}_0 \create{b}_1)\bra{0}$.
One can see that a particular choice of sign $s$ allows one to herald the states which are the closest to a conventional Bell basis.
In particular, for $s=+1$ the modified scheme operates similarly, and the deleted $30^\circ$ element effectively rotates
$\frac{1}{2} \bra{\psi^+} - \frac{\sqrt{3}}{2} \bra{\phi^-}$ to $\bra{\psi^+}$.

\section{Discussion}
We have presented a general methodology for numerical search of optimal linear optical circuits for heralded entanglement generation. We discussed its application to design the circuit for 3-GHZ state of dual-rail encoded photonic qubits. The obtained circuit has a success probability of $1/54$. Importantly, the proposed circuit does not require any feed-forward and may be used with detectors resolving up to two photons, and, to the best of our knowledge, its success probability surpasses all known results for such type of linear optical entangling gates. It is important to note that although our heralding scheme (as well as most of the others) requires minimal photon number resolution to detect the required ancillary Fock state correctly, recent progress in photon-number-resolving SNSPDs \cite{Michler2019, Berggren2019} indicates that this technology is now available. Although the ancillary states heralding successful outcomes in our schemes contain only single-photon or vacuum states in each ancillary mode, the photon number resolution is required to dismiss the terms with higher occupation numbers leading to incorrect results. At the same time, since the gate scheme is designed in such a way, that the number of ancillary photons heralding a successful result is always three and they are distributed in three distinct modes (2 in two modes for Bell states) experimentally challenging heralding on vacuum is not required.

We also identify an elementary subcircuit which enables the dual-rail encoded Bell state generation with probability $2/27$ and has the potential to be applied to other entanglement generation problems in linear optical systems. 

An important issue in real-world implementations of linear-optical circuits is the effect of loss. If the photon loss probability is uniform for all channels of the circuit, the overall effect will be just in reduction of the success probability. If, however, different channels experience different loss, fidelity of the heralded state may be compromised~\cite{Laing2017, Burgwal2017,Fldzhyan2020}. If the proposed scheme is to be realized as an integrated optical circuit, it is therefore important to design the circuit topology in such a way, that all the relevant paths are of equal length, such that the optical loss is distributed uniformly. This is, however, a general requirement for any implementation of an optical mode transforming unitary \cite{Clements}. 

The main advantage of our numerical method, as shown by these two examples, is the possibility to find simple decompositions for the required unitaries by an optimization procedure, which may be used to bring new insight to the linear optical entangling gate design.

\section{Acknowledgements}
The authors acknowledge financial support under the Russian National Technological Initiative via MSU Quantum Technology Centre and RFBR grant 19-52-80034. I.V.Dyakonov acknowledges support from RFBR grant 19-32-80020.
\bibliography{citations}
\end{document}